\documentclass[aps,pra,superscriptaddress,reprint,showpacs,10pt]{revtex4-1}
\usepackage{amsmath}
\usepackage{bm}
\usepackage{graphicx}
\usepackage{upgreek}

\begin{document}
\title{Transport of dipolar Bose-Einstein condensates in a one-dimensional optical lattice}
\author{S. K\"uhn and T. E. Judd}
\affiliation{CQ Center for Collective Quantum Phenomena and their Applications in LISA+, Physikalisches Institut, Eberhard-Karls-Universit{\"a}t T{\"u}bingen, Auf der Morgenstelle 14, D-72076 T{\"u}bingen, Germany}
\date{\today}
\begin{abstract}
We show that magnetic dipolar interactions can stabilize superfluidity in atomic gases but the dipole alignment direction required to achieve this varies, depending on whether the flow is oscillatory or continuous. If a condensate is made to oscillate through a lattice, damping of the oscillations can be reduced by aligning the dipoles perpendicular to the direction of motion. However, if a lattice is driven continuously through the condensate, superfluid behavior is best preserved when the dipoles are aligned parallel to the direction of motion. We explain these results in terms of the formation of topological excitations and tunnel barrier heights between lattice sites.
\end{abstract}
\pacs{67.85.De, 37.10.Jk, 75.10.-b}
\maketitle

\section{Introduction}
Since the development of optical lattices for cold atoms \cite{Morsch2006} there has been a wealth of research in this area, not least on transport properties. Recent developments in the cooling of atoms with strong magnetic dipolar interactions \cite{Griesmaier2005,Beaufils2008,Lu2011,Aikawa2012} have broadened perspectives further, with the possibility to explore how magnetic effects influence superflow; analogous problems in solid state physics have been a matter of debate for many years \cite{Zhu1994}. 

To date there have been a number of experimental studies on continuous and oscillatory transport of cold atoms in periodic potentials with non-magnetic atoms \cite{Denschlag2002}, including work on Bloch oscillations \cite{Anderson1998,Morsch2001}, critical velocities and dynamical instabilities \cite{Burger2001,Fallani2004}. The first experiments with dipolar gases in optical lattices are now starting to reveal information about static properties \cite{Muller2011}.
Theoretical work on gases in optical lattices has primarily focused on static phase diagrams \cite{Oosten2001} and dipolar interactions have also been considered in such cases \cite{Goral2002a}. Dynamical effects in trapped dipolar gases without a lattice potential have been studied \cite{vanBijnen2010}, such as excitation spectra \cite{Santos2003,Ronen2006a}, and anisotropic superfluidity \cite{Wilson2010,Ticknor2011}. 

However, there has been little work on the transport of dipolar gases in optical lattices \cite{Fattori2008}. In particular there has been no consideration of how the directional alignment of magnetic dipoles in such systems affects the superfluid properties. There has been theoretical work on lattice transport employing band structure calculations in one-dimension \cite{Wu2002,Machholm2003} but these are difficult to extend to many experimentally relevant situations in higher dimensions, especially when non-local interactions must be taken into account \cite{Lin2008}. In addition such approaches leave out certain details of the dynamical evolution of the cloud \cite{Brazhnyi2004}.
Saito et al.\ used a Hubbard model with a Gutzwiller Ansatz to show that dipoles aligned perpendicular to a two-dimensional lattice plane had only a small effect on dynamical instability thresholds, but other configurations were not considered.

Here we show, using dynamical simulations of the non-local Gross-Pitaevskii equation, that the directional alignment of magnetic dipoles in a Bose-Einstein condensate plays a crucial role in determining its transport properties through a shallow  one-dimensional optical lattice. Interestingly, the direction required to best stabilize the superfluidity varies depending on whether we consider oscillatory flow or continuous flow. We show this occurs because the physical processes that break down the superflow differ in the two cases. In the case of oscillatory flow, superfluidity is broken down by the formation of topological excitations. In the case of continuous flow no such excitations are observed, but dipolar interactions act to increase or decrease the tunneling barrier between lattice sites, depending on the dipole orientation.

The paper is organized as follows: in section \ref{sec:numerics} we explain the numerical methods to solve the non-local Gross-Pitaevskii equation for a dipolar gas in an optical lattice. In section \ref{sec:results} we present our results for continuous and oscillatory flow. Finally, we conclude in section \ref{sec:conclusion}.

\section{Numerical Approach}\label{sec:numerics}
In both the oscillatory and continuous cases we model the dynamics of the dipolar Bose-Einstein condensate (BEC) using the full three-dimensional Gross-Pitaevskii equation (GPE) \cite{Judd2008,Judd2010}, including the usual integral term for the dipolar potential \cite{Goral2002b}
\begin{align}
\begin{alignedat}{3}
 i\hbar\frac{\partial}{\partial t} \psi(\mathbf{r},t)= &\left[-\frac{\hbar^2}{2m}\nabla^2+ V_\mathrm{ext}(\mathbf{r}) + g|\psi(\mathbf{r},t)|^2\right.\\
& + \left.\int\mathrm{d}\mathbf{r}'\, |\psi(\mathbf{r}',t)|^2V_\mathrm{dd}(\mathbf{r}-\mathbf{r}') \right]\psi(\mathbf{r},t),
\end{alignedat}
\label{gpe_dipolar}
\end{align}
where $\psi(\mathbf{r},t)$ is the wave function at position $\mathbf{r}$ and time $t$, normalized to the total particle number $N=10^4$, $m$ is the mass of a single particle and $g=4\pi\hbar^2 a/m$ is the contact interaction strength with $a$ the $s$-wave scattering length; the other symbols have their usual meaning. The external potential
\begin{align}
V_\mathrm{ext}(\mathbf{r}) = V_0\sin^2\left(\frac{\pi x}{d}\right) + \frac{1}{2} m\left(\omega_x^2 x^2 + \omega_y^2 y^2 + \omega_z^2 z^2\right)
\end{align}
includes the optical lattice with spatial period $d$ and depth $V_0$, and the harmonic trapping potential with the trapping frequencies $\omega_x$, $\omega_y$ and $\omega_z$. The potential $V_\mathrm{dd}(\mathbf{r})$ describes the interaction between two polarized magnetic dipoles and is given by
\begin{align}
 V_\mathrm{dd}(\mathbf{r}) = \frac{\mu_0|\bm{\mu}|^2}{4\pi}\frac{1-3\cos^2(\theta)}{|\mathbf{r}|^3}
\end{align}
where $\theta$ is the angle between an external magnetic field which polarizes the dipoles and the vector $\mathbf{r}$ between the position of the two dipoles (we assume perfect alignment with the field), $\bm{\mu}$ is the dipole moment and $\mu_0$ is the vacuum permeability. We choose $^{52}\mathrm{Cr}$ atoms for our simulations, as in recent experiments \cite{Muller2011} which have $|\bm{\mu}|=6\mu_\mathrm{B}$ where $\mu_\mathrm{B}$ is the Bohr magneton and we set $a=2.0 \times 10^{-9}\:\mathrm{m}$ \footnote{The scattering length can be tuned through a wide range using Feshbach resonances. This is a typical value \cite{Lahaye2007}.}. For the optical lattice we assume $d=1.59\:\mu\mathrm{m}$ and $V_0=1.53\:E_r$ ($E_r = \hbar^2\pi^2/2md^2$ is the recoil energy). The frequencies of the harmonic trap are chosen to be $(\omega_x,\omega_y,\omega_z) = 2\pi\times(48\:\mathrm{Hz},16\:\mathrm{Hz},16\:\mathrm{Hz})$. With these parameters we obtain a system that is deep in the superfluid regime.
 
To solve Eq.\ \ref{gpe_dipolar}, we use a standard Fourier split-step method where we treat the dipolar term with the convolution theorem \cite{Goral2002b,Lahaye2008}. We prepare the numerical BEC ground state by starting with the Gaussian analytical solution for the harmonic trap \cite{Pethick2008} and ramping up the interactions and the lattice potential adiabatically. 

For each model system we look at three configurations to determine the influence of the dipole-dipole interaction (DDI). For the first configuration we neglect the DDI and look at the system with pure contact interactions which will be referred to as the Contact Configuration (this system is fictitious and is used purely for academic comparison). Secondly, we look at the system with the dipoles polarized along the $x$-direction (parallel to the direction of motion), which we call the Parallel Configuration. Due to the geometry of our model system the DDI raises the total potential energy in this case. Thirdly, we look at the system with the dipoles polarized along the $y$-direction (perpendicular to the direction of motion), henceforth referred to as the Perpendicular Configuration. In this case the DDI reduces the total potential energy. In the following section \ref{sec:results} we always depict the results for the Contact Configuration with red solid curves, and the results for the Parallel Configuration (Perpendicular Configuration) with green dashed curves (blue dotted curves).

\section{Results}\label{sec:results}
Here we present our results for oscillatory flow and continuous flow. We are primarily interested in the breakdown (or preservation) of superfluidity depending on the direction of the magnetic dipoles. Unfortunately, the superfluid fraction in our systems cannot be unambiguously defined in a useful, quantitative way. This is primarily due to the non-equilibrium and inhomogeneous nature of the system \cite{Penrose1956,Goral2002,Dodd1997,Blakie2005,Huang1992,Krumnow2011}. Naive use of the Landau criterion is further complicated by the spatially extended nature of the lattice potential that perturbs the system \cite{Pethick2008}. We therefore make use of more qualitative measures. In the oscillatory case, we consider the damping of in-trap dipole oscillations as has been done previously \cite{Saito2012}. In the continuous case, we consider the total energy of the BEC; a superfluid gas will acquire no energy from the moving lattice, therefore an increase in the BEC energy indicates a breakdown of superfluidity.

\subsection{Oscillatory flow}
\begin{figure}
	\includegraphics[width=0.9\columnwidth]{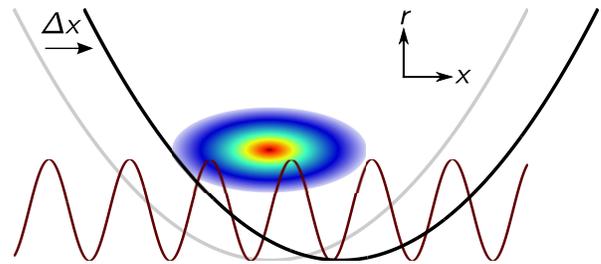}
	\caption{(Color online) Sketch of the model system for oscillatory flow. The trap is displaced a distance $\Delta x$ in the $x$-direction to trigger a center-of-mass oscillation. The colored oval represents the BEC, the gray line shows the trap before the displacement, the black line the trap after displacement and the oscillating dark red line the optical lattice. Black arrows indicate the axes with $r$ the radial direction.}
	\label{fig:falle_bewegt_gitter}
\end{figure}

\begin{figure}
	\includegraphics[width=0.9\columnwidth]{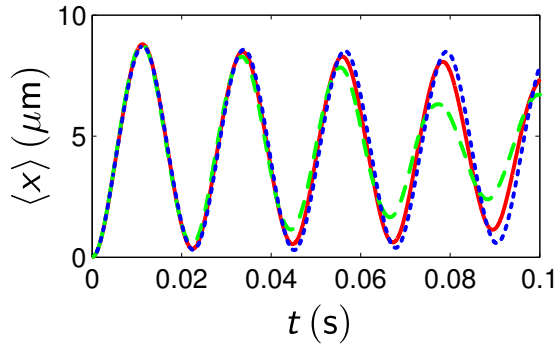}
	\caption{(Color online) Expectation value of the BEC position along the $x$-direction for the three configurations: without dipolar interactions (red solid curve), with dipoles polarized parallel to the direction of motion (green dashed curve) and with dipoles polarized perpendicular to the direction of motion (blue dotted curve).}
	\label{fig:ac_xvont}
\end{figure}

\begin{figure}
	\includegraphics[width=0.9\columnwidth]{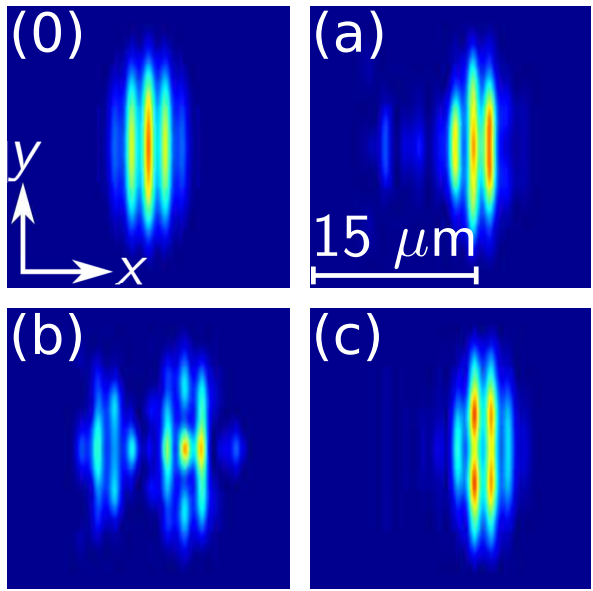}
	\caption{(Color online) Cross-section through the $z=0$ plane of the atom density distribution after $t=0.08\:\mathrm{s}$ oscillatory flow for the configuration without dipolar interactions (a), with the dipoles polarized parallel to the direction of motion (b), and with the dipoles aligned perpendicular to the direction of motion (c). For comparison we also show the profile at $t=0.0$ without dipolar interactions (0).}
	\label{fig:ac_system}
\end{figure}

To create oscillatory flow we displace the trap a distance $\Delta x$ along the $x$-direction at $t=0.0$ and trigger a center-of-mass oscillation as sketched in Fig.\ \ref{fig:falle_bewegt_gitter}. 
\noindent\noindent
Due to the harmonic trap the BEC oscillates back and forth with a velocity that varies sinusoidally (we do not fulfill the conditions for Bloch oscillations, a displacement of approximately $7\:\mu$m would be required for the condensate to reach the edge of the first Brillouin zone). 
In Fig.\ \ref{fig:ac_xvont} we show the expectation value of the BEC's position in the $x$-direction as a function of time for a trap displacement of $\Delta x = {4.5}\:\mu\mathrm{m}$ (this gives a velocity amplitude of $\omega_x \Delta x  \approx 1.4\:$mm\:s$^{-1}$). The Contact Configuration (red solid curve) shows slight damping of the center-of-mass motion during the simulation. However, the Perpendicular Configuration (blue dotted curve) shows less damping compared with the red solid curve. By contrast, the Parallel Configuration (green dashed curve) shows stronger damping. 
\noindent
To further investigate the system we consider snapshots of the density profile following several oscillations in the trap. Figure \ref{fig:ac_system} shows the disruption of the BEC density profile as a result of oscillatory flow. 
\noindent
Panel (0) shows the profile at the start of the simulation before flow has begun ($t=0.0$). We use the Contact Configuration here but differences due to the influence of the DDI are barely discernible at this point. Panels (a-c) show the density profile after $0.08\:\mathrm{s}$ oscillatory flow (around 4 oscillation periods). Panel (a) shows the Contact Configuration; we see that a limited amount of disruption has occurred, compared with the situation at $t=0.0$, the cloud acquiring a small ``tail'' to the left. By contrast, panel (b) shows the Parallel Configuration. We see significantly more disruption in this case, including a large density minimum in the center, and radial topological excitations. These topological excitations typically consist of solitons and vortices and form as a result of the direction reversal of the BEC's motion \cite{Scott2004} which pushes atoms into the radial direction $r$; this nonlinear mixing explains the damping behavior in Fig.\ \ref{fig:ac_xvont}. Panel (c) shows the profile for the Perpendicular Configuration. In this case, the ``tail'' is smaller and the cloud has largely retained its form, compared with the starting profile.

The reason we see less damping for the Perpendicular Configuration is because the DDI creates potential wells that restrict motion in the radial direction and therefore the cloud is more stable against motion away from the primary direction of flow. By contrast the Parallel Configuration amplifies the creation of topological excitations because the DDI raises the potential energy, and thereby the level of instability.

\subsection{Continuous flow}
\begin{figure}
	\includegraphics[width=0.9\columnwidth]{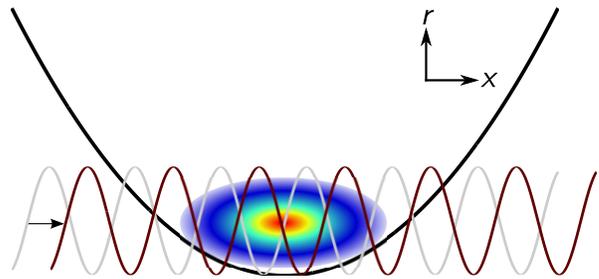}
	\caption{(Color online) Sketch of the model system for continuous flow. The colored oval represents the BEC. The optical lattice is dragged along the $x$-axis through the cloud at constant velocity. The black line shows the harmonic trap, the oscillating gray curve shows the optical lattice at $t=0.0$ and the oscillating dark red curve shows the shifted lattice. Black arrows indicate the axes with $r$ the radial direction.}
	\label{fig:falle_gitter_bewegt}
\end{figure}

\begin{figure}
	\includegraphics[width=1.0\columnwidth]{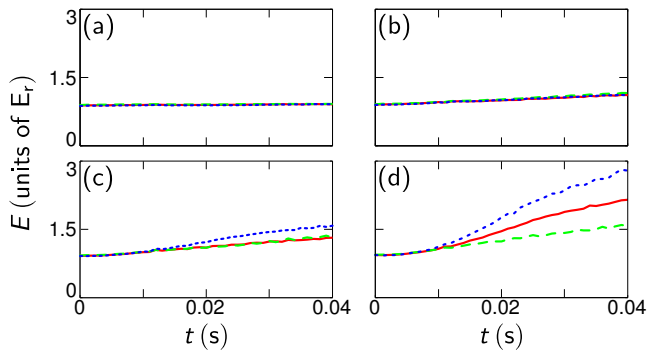}
	\caption{(Color online) Total energy per atom as a function of time for lattice velocities $0.75\:\mathrm{mm\,s}^{-1}$ (a), $0.98\:\mathrm{mm\,s}^{-1}$ (b), $1.06\:\mathrm{mm\,s}^{-1}$ (c) and $1.13\:\mathrm{mm\,s}^{-1}$ (d). The red solid curves show the configuration without dipolar interactions, the green dashed curves the configuration with the dipoles aligned parallel to the direction of motion and the blue dotted curves the configuration with the dipoles aligned perpendicular to the direction of motion.}
	\label{fig:dc_gesamt}
\end{figure}

In our second system we drag the optical lattice through the BEC at constant velocity $v$, as sketched in Fig.\ \ref{fig:falle_gitter_bewegt}. 
\noindent
Figure \ref{fig:dc_gesamt} shows the total energy per atom as a function of time for different $v$. We see that below $v\approx 1.00\:\mathrm{mm\,s}^{-1}$, little energy is transferred to the BEC [Panel (a)]. However, around this velocity, the three configurations begin to gain energy [Panel (b)]. Initially, this energy gain is similar in all three cases but for higher velocities, the curves start to diverge from one another [Panel (c) and Panel (d)]. The Perpendicular Configuration (blue dotted curve) shows a greater increase in energy than the others whereas the Parallel Configuration (green dashed curve) has the smallest energy increase. The Contact Configuration (red solid curve) lies between the other two configurations. 
\noindent
We see that the increase in energy only becomes significant after $\sim 10\:$ms. For this reason, these instability effects are unlikely to have played a significant role in the oscillatory case, even though the velocities were above the required thresholds; the condensate spent too little time above the thresholds for the atoms to respond.

\begin{figure}
	\includegraphics[width=0.9\columnwidth]{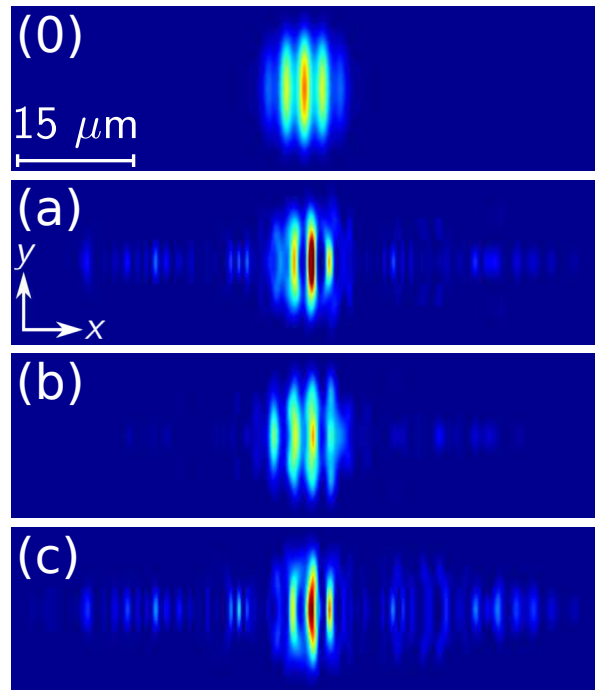}
	\caption{(Color online) Cross-section through the $z=0$ plane of the atom density distribution after $t=0.04\:\mathrm{s}$ continuous flow for the configuration without dipolar interactions (a), with the dipoles polarized parallel to the direction of motion (b) and with the dipoles aligned perpendicular to the direction of motion (c). The lattice velocity is $v=1.13\:\mathrm{mm\,s}^{-1}$. For comparison we also show the profile at $t=0.0$ without dipolar interactions (0).}
	\label{fig:dc_system}
\end{figure}

As in the case of oscillatory flow, we explore this behavior further by examining the cloud density profiles. Figure \ref{fig:dc_system} shows the density profiles of the gas after $0.04\:\mathrm{s}$ flow time with $v=1.13\:\mathrm{mm\,s}^{-1}$. 
\noindent
Panel (0) shows the starting profile at $t=0.0$ as before. Panel (a) shows the profile for the Contact Configuration. We see that significant numbers of atoms have been displaced from their original positions in the center. We are at the edge of the zone for dynamical instability where perturbations in both the $+x$ and $-x$ directions can grow exponentially in time. The density of these ``wings'' is noticeably less for the Parallel Configuration [Panel (b)]. By contrast, the Perpendicular Configuration shows increased density in the wings [Panel (c)]. Unlike the oscillatory flow, we see no large density minima and little evidence of radial topological excitations.

\begin{figure}
	\includegraphics[width=0.9\columnwidth]{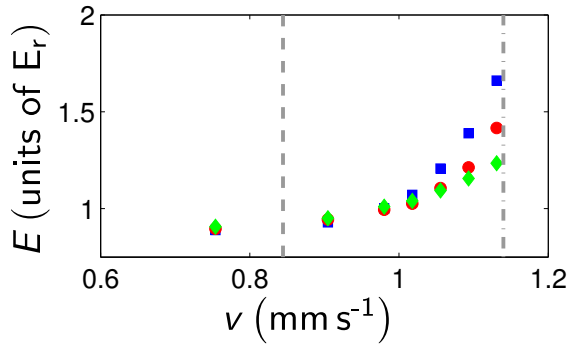}
	\caption{(Color online) Total energy per atom at $t=0.02\:\mathrm{s}$ as a function of $v$ for the three configurations: without dipolar interactions (red dots), with the dipoles polarized parallel to the direction of motion (green diamonds) and with the dipoles polarized perpendicular to the direction of motion (blue squares). Vertical gray lines show the thresholds for the energetic instability (dashed curve) and the dynamical instability (dot-dashed cure).}
	\label{fig:dc_evonv}
\end{figure}

To summarize these results, we show the total energy per atom as a function of the lattice velocity in Fig.\ \ref{fig:dc_evonv}. The values were taken after $t=0.02\:\mathrm{s}$ which is well above the correlation time of the BEC. The energy increase is noticeably more severe for the Perpendicular Configuration than the Parallel Configuration. For comparison, instability thresholds have been calculated by determining the lowest energy band and checking its stability against perturbations following reference \cite{Machholm2003} with pure contact interactions \footnote{We have not performed these calculations including the dipolar interactions for the numerical reasons already discussed.}. The thresholds for energetic and dynamical instability from the band structure calculations correspond well to all three of our configurations as expected \cite{Saito2012}.
\noindent

That the Perpendicular Configuration can stabilize superflow for oscillatory motion but destabilize it for continuous flow (vice-versa for the Parallel Configuration) may initially seem paradoxical. However, we have observed that the superflow is not broken down by topological excitations in the continuous case, as it is in the oscillatory case. We therefore turn our attention to tunneling between lattice sites by considering the effective lattice potential $V_\mathrm{eff}(\mathbf{r})$ which is the sum of the optical lattice, the contact interaction and the dipolar interaction. If we compare the depth of $V_\mathrm{eff}(\mathbf{r})$ at $x=0$ for the Perpendicular Configuration with the Contact Configuration, we find that the DDI increases the effective lattice depth by about $3.0\%$, making it harder for atoms to tunnel. Conversely, the Parallel Configuration lowers the depth of the effective potential by about $3.6\%$, effectively increasing the hopping coefficient. Consequently, the reduction (increase) of the tunnel barrier facilitates (hinders) the motion of the BEC through the lattice, thereby explaining the results in Figs.\ \ref{fig:dc_gesamt} - \ref{fig:dc_evonv}.

\section{Conclusions}\label{sec:conclusion}
In conclusion, we have studied superfluid transport of magnetic dipolar Bose-Einstein condensates in one-dimensional optical lattices. We have shown that the dipolar interactions can be used to stabilize the superflow in such systems but the atomic dipoles must be aligned in different directions to achieve this, depending on whether the flow of the gas is oscillatory (``alternating current'') or continuous (``direct current''). In spite of the apparently paradoxical nature of this result, we showed it could be simply explained by the dipoles' effect on topological excitations and optical lattice barrier height. 
The direction reversal in the alternating case is the key difference that gives rise to different physics.

It is likely that the effects we studied here will be even more pronounced for cold atoms with stronger magnetic dipole moments such as Erbium and Dysprosium, assuming that possible instabilities can be controlled. There are further questions relating to two- and three-dimensional lattices; interesting phenomena may occur in deeper lattices. In addition, there is the possibility of considering finite-temperature gases and studying the effects of magnetic dipoles on the superfluid critical temperature.

We thank T. Pfau and A. Griesmaier for helpful discussions and cooperations within the SFB TR21. We also thank R. Scott and A. M. Martin for discussions, and BW-Grid computing resources.


\begin{thebibliography}{40}%
\makeatletter
\providecommand \@ifxundefined [1]{%
 \@ifx{#1\undefined}
}%
\providecommand \@ifnum [1]{%
 \ifnum #1\expandafter \@firstoftwo
 \else \expandafter \@secondoftwo
 \fi
}%
\providecommand \@ifx [1]{%
 \ifx #1\expandafter \@firstoftwo
 \else \expandafter \@secondoftwo
 \fi
}%
\providecommand \natexlab [1]{#1}%
\providecommand \enquote  [1]{``#1''}%
\providecommand \bibnamefont  [1]{#1}%
\providecommand \bibfnamefont [1]{#1}%
\providecommand \citenamefont [1]{#1}%
\providecommand \href@noop [0]{\@secondoftwo}%
\providecommand \href [0]{\begingroup \@sanitize@url \@href}%
\providecommand \@href[1]{\@@startlink{#1}\@@href}%
\providecommand \@@href[1]{\endgroup#1\@@endlink}%
\providecommand \@sanitize@url [0]{\catcode `\\12\catcode `\$12\catcode
  `\&12\catcode `\#12\catcode `\^12\catcode `\_12\catcode `\%12\relax}%
\providecommand \@@startlink[1]{}%
\providecommand \@@endlink[0]{}%
\providecommand \url  [0]{\begingroup\@sanitize@url \@url }%
\providecommand \@url [1]{\endgroup\@href {#1}{\urlprefix }}%
\providecommand \urlprefix  [0]{URL }%
\providecommand \Eprint [0]{\href }%
\providecommand \doibase [0]{http://dx.doi.org/}%
\providecommand \selectlanguage [0]{\@gobble}%
\providecommand \bibinfo  [0]{\@secondoftwo}%
\providecommand \bibfield  [0]{\@secondoftwo}%
\providecommand \translation [1]{[#1]}%
\providecommand \BibitemOpen [0]{}%
\providecommand \bibitemStop [0]{}%
\providecommand \bibitemNoStop [0]{.\EOS\space}%
\providecommand \EOS [0]{\spacefactor3000\relax}%
\providecommand \BibitemShut  [1]{\csname bibitem#1\endcsname}%
\let\auto@bib@innerbib\@empty
%</preamble>
\bibitem [{\citenamefont {Morsch}\ and\ \citenamefont
  {Oberthaler}(2006)}]{Morsch2006}%
  \BibitemOpen
  \bibfield  {author} {\bibinfo {author} {\bibfnamefont {O.}~\bibnamefont
  {Morsch}}\ and\ \bibinfo {author} {\bibfnamefont {M.}~\bibnamefont
  {Oberthaler}},\ }\href {\doibase 10.1103/RevModPhys.78.179} {\bibfield
  {journal} {\bibinfo  {journal} {Rev. Mod. Phys.}\ }\textbf {\bibinfo {volume}
  {78}},\ \bibinfo {pages} {179} (\bibinfo {year} {2006})}\BibitemShut
  {NoStop}%
\bibitem [{\citenamefont {Griesmaier}\ \emph {et~al.}(2005)\citenamefont
  {Griesmaier}, \citenamefont {Werner}, \citenamefont {Hensler}, \citenamefont
  {Stuhler},\ and\ \citenamefont {Pfau}}]{Griesmaier2005}%
  \BibitemOpen
  \bibfield  {author} {\bibinfo {author} {\bibfnamefont {A.}~\bibnamefont
  {Griesmaier}}, \bibinfo {author} {\bibfnamefont {J.}~\bibnamefont {Werner}},
  \bibinfo {author} {\bibfnamefont {S.}~\bibnamefont {Hensler}}, \bibinfo
  {author} {\bibfnamefont {J.}~\bibnamefont {Stuhler}}, \ and\ \bibinfo
  {author} {\bibfnamefont {T.}~\bibnamefont {Pfau}},\ }\href {\doibase
  10.1103/PhysRevLett.94.160401} {\bibfield  {journal} {\bibinfo  {journal}
  {Phys. Rev. Lett.}\ }\textbf {\bibinfo {volume} {94}},\ \bibinfo {pages}
  {160401} (\bibinfo {year} {2005})}\BibitemShut {NoStop}%
\bibitem [{\citenamefont {Beaufils}\ \emph {et~al.}(2008)\citenamefont
  {Beaufils}, \citenamefont {Chicireanu}, \citenamefont {Zanon}, \citenamefont
  {Laburthe-Tolra}, \citenamefont {Mar\'echal}, \citenamefont {Vernac},
  \citenamefont {Keller},\ and\ \citenamefont {Gorceix}}]{Beaufils2008}%
  \BibitemOpen
  \bibfield  {author} {\bibinfo {author} {\bibfnamefont {Q.}~\bibnamefont
  {Beaufils}}, \bibinfo {author} {\bibfnamefont {R.}~\bibnamefont
  {Chicireanu}}, \bibinfo {author} {\bibfnamefont {T.}~\bibnamefont {Zanon}},
  \bibinfo {author} {\bibfnamefont {B.}~\bibnamefont {Laburthe-Tolra}},
  \bibinfo {author} {\bibfnamefont {E.}~\bibnamefont {Mar\'echal}}, \bibinfo
  {author} {\bibfnamefont {L.}~\bibnamefont {Vernac}}, \bibinfo {author}
  {\bibfnamefont {J.-C.}\ \bibnamefont {Keller}}, \ and\ \bibinfo {author}
  {\bibfnamefont {O.}~\bibnamefont {Gorceix}},\ }\href {\doibase
  10.1103/PhysRevA.77.061601} {\bibfield  {journal} {\bibinfo  {journal} {Phys.
  Rev. A}\ }\textbf {\bibinfo {volume} {77}},\ \bibinfo {pages} {061601}
  (\bibinfo {year} {2008})}\BibitemShut {NoStop}%
\bibitem [{\citenamefont {Lu}\ \emph {et~al.}(2011)\citenamefont {Lu},
  \citenamefont {Burdick}, \citenamefont {Youn},\ and\ \citenamefont
  {Lev}}]{Lu2011}%
  \BibitemOpen
  \bibfield  {author} {\bibinfo {author} {\bibfnamefont {M.}~\bibnamefont
  {Lu}}, \bibinfo {author} {\bibfnamefont {N.~Q.}\ \bibnamefont {Burdick}},
  \bibinfo {author} {\bibfnamefont {S.~H.}\ \bibnamefont {Youn}}, \ and\
  \bibinfo {author} {\bibfnamefont {B.~L.}\ \bibnamefont {Lev}},\ }\href
  {\doibase 10.1103/PhysRevLett.107.190401} {\bibfield  {journal} {\bibinfo
  {journal} {Phys. Rev. Lett.}\ }\textbf {\bibinfo {volume} {107}},\ \bibinfo
  {pages} {190401} (\bibinfo {year} {2011})}\BibitemShut {NoStop}%
\bibitem [{\citenamefont {Aikawa}\ \emph {et~al.}(2012)\citenamefont {Aikawa},
  \citenamefont {Frisch}, \citenamefont {Mark}, \citenamefont {Baier},
  \citenamefont {Rietzler}, \citenamefont {Grimm},\ and\ \citenamefont
  {Ferlaino}}]{Aikawa2012}%
  \BibitemOpen
  \bibfield  {author} {\bibinfo {author} {\bibfnamefont {K.}~\bibnamefont
  {Aikawa}}, \bibinfo {author} {\bibfnamefont {A.}~\bibnamefont {Frisch}},
  \bibinfo {author} {\bibfnamefont {M.}~\bibnamefont {Mark}}, \bibinfo {author}
  {\bibfnamefont {S.}~\bibnamefont {Baier}}, \bibinfo {author} {\bibfnamefont
  {A.}~\bibnamefont {Rietzler}}, \bibinfo {author} {\bibfnamefont
  {R.}~\bibnamefont {Grimm}}, \ and\ \bibinfo {author} {\bibfnamefont
  {F.}~\bibnamefont {Ferlaino}},\ }\href {\doibase
  10.1103/PhysRevLett.108.210401} {\bibfield  {journal} {\bibinfo  {journal}
  {Phys. Rev. Lett.}\ }\textbf {\bibinfo {volume} {108}},\ \bibinfo {pages}
  {210401} (\bibinfo {year} {2012})}\BibitemShut {NoStop}%
\bibitem [{\citenamefont {Zhu}\ \emph {et~al.}(1994)\citenamefont {Zhu},
  \citenamefont {Kang}, \citenamefont {Lee},\ and\ \citenamefont
  {Chen}}]{Zhu1994}%
  \BibitemOpen
  \bibfield  {author} {\bibinfo {author} {\bibfnamefont {W.}~\bibnamefont
  {Zhu}}, \bibinfo {author} {\bibfnamefont {H.}~\bibnamefont {Kang}}, \bibinfo
  {author} {\bibfnamefont {Y.~C.}\ \bibnamefont {Lee}}, \ and\ \bibinfo
  {author} {\bibfnamefont {J.~C.}\ \bibnamefont {Chen}},\ }\href {\doibase
  10.1103/PhysRevB.50.10302} {\bibfield  {journal} {\bibinfo  {journal} {Phys.
  Rev. B}\ }\textbf {\bibinfo {volume} {50}},\ \bibinfo {pages} {10302}
  (\bibinfo {year} {1994})}\BibitemShut {NoStop}%
\bibitem [{\citenamefont {Denschlag}\ \emph {et~al.}(2002)\citenamefont
  {Denschlag}, \citenamefont {Simsarian}, \citenamefont {H\"affner},
  \citenamefont {McKenzie}, \citenamefont {Browaeys}, \citenamefont {Cho},
  \citenamefont {Helmerson}, \citenamefont {Rolston},\ and\ \citenamefont
  {Phillips}}]{Denschlag2002}%
  \BibitemOpen
  \bibfield  {author} {\bibinfo {author} {\bibfnamefont {J.~H.}\ \bibnamefont
  {Denschlag}}, \bibinfo {author} {\bibfnamefont {J.~E.}\ \bibnamefont
  {Simsarian}}, \bibinfo {author} {\bibfnamefont {H.}~\bibnamefont
  {H\"affner}}, \bibinfo {author} {\bibfnamefont {C.}~\bibnamefont {McKenzie}},
  \bibinfo {author} {\bibfnamefont {A.}~\bibnamefont {Browaeys}}, \bibinfo
  {author} {\bibfnamefont {D.}~\bibnamefont {Cho}}, \bibinfo {author}
  {\bibfnamefont {K.}~\bibnamefont {Helmerson}}, \bibinfo {author}
  {\bibfnamefont {S.~L.}\ \bibnamefont {Rolston}}, \ and\ \bibinfo {author}
  {\bibfnamefont {W.~D.}\ \bibnamefont {Phillips}},\ }\href {\doibase
  10.1088/0953-4075/35/14/307} {\bibfield  {journal} {\bibinfo  {journal} {J.
  Phys. B}\ }\textbf {\bibinfo {volume} {35}},\ \bibinfo {pages} {3095}
  (\bibinfo {year} {2002})}\BibitemShut {NoStop}%
\bibitem [{\citenamefont {Anderson}\ and\ \citenamefont
  {Kasevich}(1998)}]{Anderson1998}%
  \BibitemOpen
  \bibfield  {author} {\bibinfo {author} {\bibfnamefont {B.~P.}\ \bibnamefont
  {Anderson}}\ and\ \bibinfo {author} {\bibfnamefont {M.~A.}\ \bibnamefont
  {Kasevich}},\ }\href {\doibase 10.1126/science.282.5394.1686} {\bibfield
  {journal} {\bibinfo  {journal} {Science}\ }\textbf {\bibinfo {volume}
  {282}},\ \bibinfo {pages} {1686} (\bibinfo {year} {1998})}\BibitemShut
  {NoStop}%
\bibitem [{\citenamefont {Morsch}\ \emph {et~al.}(2001)\citenamefont {Morsch},
  \citenamefont {M\"uller}, \citenamefont {Cristiani}, \citenamefont
  {Ciampini},\ and\ \citenamefont {Arimondo}}]{Morsch2001}%
  \BibitemOpen
  \bibfield  {author} {\bibinfo {author} {\bibfnamefont {O.}~\bibnamefont
  {Morsch}}, \bibinfo {author} {\bibfnamefont {J.~H.}\ \bibnamefont
  {M\"uller}}, \bibinfo {author} {\bibfnamefont {M.}~\bibnamefont {Cristiani}},
  \bibinfo {author} {\bibfnamefont {D.}~\bibnamefont {Ciampini}}, \ and\
  \bibinfo {author} {\bibfnamefont {E.}~\bibnamefont {Arimondo}},\ }\href
  {\doibase 10.1103/PhysRevLett.87.140402} {\bibfield  {journal} {\bibinfo
  {journal} {Phys. Rev. Lett.}\ }\textbf {\bibinfo {volume} {87}},\ \bibinfo
  {pages} {140402} (\bibinfo {year} {2001})}\BibitemShut {NoStop}%
\bibitem [{\citenamefont {Burger}\ \emph {et~al.}(2001)\citenamefont {Burger},
  \citenamefont {Cataliotti}, \citenamefont {Fort}, \citenamefont {Minardi},
  \citenamefont {Inguscio}, \citenamefont {Chiofalo},\ and\ \citenamefont
  {Tosi}}]{Burger2001}%
  \BibitemOpen
  \bibfield  {author} {\bibinfo {author} {\bibfnamefont {S.}~\bibnamefont
  {Burger}}, \bibinfo {author} {\bibfnamefont {F.~S.}\ \bibnamefont
  {Cataliotti}}, \bibinfo {author} {\bibfnamefont {C.}~\bibnamefont {Fort}},
  \bibinfo {author} {\bibfnamefont {F.}~\bibnamefont {Minardi}}, \bibinfo
  {author} {\bibfnamefont {M.}~\bibnamefont {Inguscio}}, \bibinfo {author}
  {\bibfnamefont {M.~L.}\ \bibnamefont {Chiofalo}}, \ and\ \bibinfo {author}
  {\bibfnamefont {M.~P.}\ \bibnamefont {Tosi}},\ }\href {\doibase
  10.1103/PhysRevLett.86.4447} {\bibfield  {journal} {\bibinfo  {journal}
  {Phys. Rev. Lett.}\ }\textbf {\bibinfo {volume} {86}},\ \bibinfo {pages}
  {4447} (\bibinfo {year} {2001})}\BibitemShut {NoStop}%
\bibitem [{\citenamefont {Fallani}\ \emph {et~al.}(2004)\citenamefont
  {Fallani}, \citenamefont {De~Sarlo}, \citenamefont {Lye}, \citenamefont
  {Modugno}, \citenamefont {Saers}, \citenamefont {Fort},\ and\ \citenamefont
  {Inguscio}}]{Fallani2004}%
  \BibitemOpen
  \bibfield  {author} {\bibinfo {author} {\bibfnamefont {L.}~\bibnamefont
  {Fallani}}, \bibinfo {author} {\bibfnamefont {L.}~\bibnamefont {De~Sarlo}},
  \bibinfo {author} {\bibfnamefont {J.~E.}\ \bibnamefont {Lye}}, \bibinfo
  {author} {\bibfnamefont {M.}~\bibnamefont {Modugno}}, \bibinfo {author}
  {\bibfnamefont {R.}~\bibnamefont {Saers}}, \bibinfo {author} {\bibfnamefont
  {C.}~\bibnamefont {Fort}}, \ and\ \bibinfo {author} {\bibfnamefont
  {M.}~\bibnamefont {Inguscio}},\ }\href {\doibase
  10.1103/PhysRevLett.93.140406} {\bibfield  {journal} {\bibinfo  {journal}
  {Phys. Rev. Lett.}\ }\textbf {\bibinfo {volume} {93}},\ \bibinfo {pages}
  {140406} (\bibinfo {year} {2004})}\BibitemShut {NoStop}%
\bibitem [{\citenamefont {M\"uller}\ \emph {et~al.}(2011)\citenamefont
  {M\"uller}, \citenamefont {Billy}, \citenamefont {Henn}, \citenamefont
  {Kadau}, \citenamefont {Griesmaier}, \citenamefont {Jona-Lasinio},
  \citenamefont {Santos},\ and\ \citenamefont {Pfau}}]{Muller2011}%
  \BibitemOpen
  \bibfield  {author} {\bibinfo {author} {\bibfnamefont {S.}~\bibnamefont
  {M\"uller}}, \bibinfo {author} {\bibfnamefont {J.}~\bibnamefont {Billy}},
  \bibinfo {author} {\bibfnamefont {E.~A.~L.}\ \bibnamefont {Henn}}, \bibinfo
  {author} {\bibfnamefont {H.}~\bibnamefont {Kadau}}, \bibinfo {author}
  {\bibfnamefont {A.}~\bibnamefont {Griesmaier}}, \bibinfo {author}
  {\bibfnamefont {M.}~\bibnamefont {Jona-Lasinio}}, \bibinfo {author}
  {\bibfnamefont {L.}~\bibnamefont {Santos}}, \ and\ \bibinfo {author}
  {\bibfnamefont {T.}~\bibnamefont {Pfau}},\ }\href {\doibase
  10.1103/PhysRevA.84.053601} {\bibfield  {journal} {\bibinfo  {journal} {Phys.
  Rev. A}\ }\textbf {\bibinfo {volume} {84}},\ \bibinfo {pages} {053601}
  (\bibinfo {year} {2011})}\BibitemShut {NoStop}%
\bibitem [{\citenamefont {van Oosten}\ \emph {et~al.}(2001)\citenamefont {van
  Oosten}, \citenamefont {van~der Straten},\ and\ \citenamefont
  {Stoof}}]{Oosten2001}%
  \BibitemOpen
  \bibfield  {author} {\bibinfo {author} {\bibfnamefont {D.}~\bibnamefont {van
  Oosten}}, \bibinfo {author} {\bibfnamefont {P.}~\bibnamefont {van~der
  Straten}}, \ and\ \bibinfo {author} {\bibfnamefont {H.~T.~C.}\ \bibnamefont
  {Stoof}},\ }\href {\doibase 10.1103/PhysRevA.63.053601} {\bibfield  {journal}
  {\bibinfo  {journal} {Phys. Rev. A}\ }\textbf {\bibinfo {volume} {63}},\
  \bibinfo {pages} {053601} (\bibinfo {year} {2001})}\BibitemShut {NoStop}%
\bibitem [{\citenamefont {G\'oral}\ \emph
  {et~al.}(2002{\natexlab{a}})\citenamefont {G\'oral}, \citenamefont {Santos},\
  and\ \citenamefont {Lewenstein}}]{Goral2002a}%
  \BibitemOpen
  \bibfield  {author} {\bibinfo {author} {\bibfnamefont {K.}~\bibnamefont
  {G\'oral}}, \bibinfo {author} {\bibfnamefont {L.}~\bibnamefont {Santos}}, \
  and\ \bibinfo {author} {\bibfnamefont {M.}~\bibnamefont {Lewenstein}},\
  }\href {\doibase 10.1103/PhysRevLett.88.170406} {\bibfield  {journal}
  {\bibinfo  {journal} {Phys. Rev. Lett.}\ }\textbf {\bibinfo {volume} {88}},\
  \bibinfo {pages} {170406} (\bibinfo {year} {2002}{\natexlab{a}})}\BibitemShut
  {NoStop}%
\bibitem [{\citenamefont {van Bijnen}\ \emph {et~al.}(2010)\citenamefont {van
  Bijnen}, \citenamefont {Parker}, \citenamefont {Kokkelmans}, \citenamefont
  {Martin},\ and\ \citenamefont {O'Dell}}]{vanBijnen2010}%
  \BibitemOpen
  \bibfield  {author} {\bibinfo {author} {\bibfnamefont {R.~M.~W.}\
  \bibnamefont {van Bijnen}}, \bibinfo {author} {\bibfnamefont {N.~G.}\
  \bibnamefont {Parker}}, \bibinfo {author} {\bibfnamefont {S.~J. J. M.~F.}\
  \bibnamefont {Kokkelmans}}, \bibinfo {author} {\bibfnamefont {A.~M.}\
  \bibnamefont {Martin}}, \ and\ \bibinfo {author} {\bibfnamefont {D.~H.~J.}\
  \bibnamefont {O'Dell}},\ }\href {\doibase 10.1103/PhysRevA.82.033612}
  {\bibfield  {journal} {\bibinfo  {journal} {Phys. Rev. A}\ }\textbf {\bibinfo
  {volume} {82}},\ \bibinfo {pages} {033612} (\bibinfo {year}
  {2010})}\BibitemShut {NoStop}%
\bibitem [{\citenamefont {Santos}\ \emph {et~al.}(2003)\citenamefont {Santos},
  \citenamefont {Shlyapnikov},\ and\ \citenamefont {Lewenstein}}]{Santos2003}%
  \BibitemOpen
  \bibfield  {author} {\bibinfo {author} {\bibfnamefont {L.}~\bibnamefont
  {Santos}}, \bibinfo {author} {\bibfnamefont {G.~V.}\ \bibnamefont
  {Shlyapnikov}}, \ and\ \bibinfo {author} {\bibfnamefont {M.}~\bibnamefont
  {Lewenstein}},\ }\href {\doibase 10.1103/PhysRevLett.90.250403} {\bibfield
  {journal} {\bibinfo  {journal} {Phys. Rev. Lett.}\ }\textbf {\bibinfo
  {volume} {90}},\ \bibinfo {pages} {250403} (\bibinfo {year}
  {2003})}\BibitemShut {NoStop}%
\bibitem [{\citenamefont {Ronen}\ \emph {et~al.}(2006)\citenamefont {Ronen},
  \citenamefont {Bortolotti},\ and\ \citenamefont {Bohn}}]{Ronen2006a}%
  \BibitemOpen
  \bibfield  {author} {\bibinfo {author} {\bibfnamefont {S.}~\bibnamefont
  {Ronen}}, \bibinfo {author} {\bibfnamefont {D.~C.~E.}\ \bibnamefont
  {Bortolotti}}, \ and\ \bibinfo {author} {\bibfnamefont {J.~L.}\ \bibnamefont
  {Bohn}},\ }\href {\doibase 10.1103/PhysRevA.74.013623} {\bibfield  {journal}
  {\bibinfo  {journal} {Phys. Rev. A}\ }\textbf {\bibinfo {volume} {74}},\
  \bibinfo {pages} {013623} (\bibinfo {year} {2006})}\BibitemShut {NoStop}%
\bibitem [{\citenamefont {Wilson}\ \emph {et~al.}(2010)\citenamefont {Wilson},
  \citenamefont {Ronen},\ and\ \citenamefont {Bohn}}]{Wilson2010}%
  \BibitemOpen
  \bibfield  {author} {\bibinfo {author} {\bibfnamefont {R.~M.}\ \bibnamefont
  {Wilson}}, \bibinfo {author} {\bibfnamefont {S.}~\bibnamefont {Ronen}}, \
  and\ \bibinfo {author} {\bibfnamefont {J.~L.}\ \bibnamefont {Bohn}},\ }\href
  {\doibase 10.1103/PhysRevLett.104.094501} {\bibfield  {journal} {\bibinfo
  {journal} {Phys. Rev. Lett.}\ }\textbf {\bibinfo {volume} {104}},\ \bibinfo
  {pages} {094501} (\bibinfo {year} {2010})}\BibitemShut {NoStop}%
\bibitem [{\citenamefont {Ticknor}\ \emph {et~al.}(2011)\citenamefont
  {Ticknor}, \citenamefont {Wilson},\ and\ \citenamefont {Bohn}}]{Ticknor2011}%
  \BibitemOpen
  \bibfield  {author} {\bibinfo {author} {\bibfnamefont {C.}~\bibnamefont
  {Ticknor}}, \bibinfo {author} {\bibfnamefont {R.~M.}\ \bibnamefont {Wilson}},
  \ and\ \bibinfo {author} {\bibfnamefont {J.~L.}\ \bibnamefont {Bohn}},\
  }\href {\doibase 10.1103/PhysRevLett.106.065301} {\bibfield  {journal}
  {\bibinfo  {journal} {Phys. Rev. Lett.}\ }\textbf {\bibinfo {volume} {106}},\
  \bibinfo {pages} {065301} (\bibinfo {year} {2011})}\BibitemShut {NoStop}%
\bibitem [{\citenamefont {Fattori}\ \emph {et~al.}(2008)\citenamefont
  {Fattori}, \citenamefont {Roati}, \citenamefont {Deissler}, \citenamefont
  {D'Errico}, \citenamefont {Zaccanti}, \citenamefont {Jona-Lasinio},
  \citenamefont {Santos}, \citenamefont {Inguscio},\ and\ \citenamefont
  {Modugno}}]{Fattori2008}%
  \BibitemOpen
  \bibfield  {author} {\bibinfo {author} {\bibfnamefont {M.}~\bibnamefont
  {Fattori}}, \bibinfo {author} {\bibfnamefont {G.}~\bibnamefont {Roati}},
  \bibinfo {author} {\bibfnamefont {B.}~\bibnamefont {Deissler}}, \bibinfo
  {author} {\bibfnamefont {C.}~\bibnamefont {D'Errico}}, \bibinfo {author}
  {\bibfnamefont {M.}~\bibnamefont {Zaccanti}}, \bibinfo {author}
  {\bibfnamefont {M.}~\bibnamefont {Jona-Lasinio}}, \bibinfo {author}
  {\bibfnamefont {L.}~\bibnamefont {Santos}}, \bibinfo {author} {\bibfnamefont
  {M.}~\bibnamefont {Inguscio}}, \ and\ \bibinfo {author} {\bibfnamefont
  {G.}~\bibnamefont {Modugno}},\ }\href {\doibase
  10.1103/PhysRevLett.101.190405} {\bibfield  {journal} {\bibinfo  {journal}
  {Phys. Rev. Lett.}\ }\textbf {\bibinfo {volume} {101}},\ \bibinfo {pages}
  {190405} (\bibinfo {year} {2008})}\BibitemShut {NoStop}%
\bibitem [{\citenamefont {Wu}\ \emph {et~al.}(2002)\citenamefont {Wu},
  \citenamefont {Diener},\ and\ \citenamefont {Niu}}]{Wu2002}%
  \BibitemOpen
  \bibfield  {author} {\bibinfo {author} {\bibfnamefont {B.}~\bibnamefont
  {Wu}}, \bibinfo {author} {\bibfnamefont {R.~B.}\ \bibnamefont {Diener}}, \
  and\ \bibinfo {author} {\bibfnamefont {Q.}~\bibnamefont {Niu}},\ }\href
  {\doibase 10.1103/PhysRevA.65.025601} {\bibfield  {journal} {\bibinfo
  {journal} {Phys. Rev. A}\ }\textbf {\bibinfo {volume} {65}},\ \bibinfo
  {pages} {025601} (\bibinfo {year} {2002})}\BibitemShut {NoStop}%
\bibitem [{\citenamefont {Machholm}\ \emph {et~al.}(2003)\citenamefont
  {Machholm}, \citenamefont {Pethick},\ and\ \citenamefont
  {Smith}}]{Machholm2003}%
  \BibitemOpen
  \bibfield  {author} {\bibinfo {author} {\bibfnamefont {M.}~\bibnamefont
  {Machholm}}, \bibinfo {author} {\bibfnamefont {C.~J.}\ \bibnamefont
  {Pethick}}, \ and\ \bibinfo {author} {\bibfnamefont {H.}~\bibnamefont
  {Smith}},\ }\href {\doibase 10.1103/PhysRevA.67.053613} {\bibfield  {journal}
  {\bibinfo  {journal} {Phys. Rev. A}\ }\textbf {\bibinfo {volume} {67}},\
  \bibinfo {pages} {053613} (\bibinfo {year} {2003})}\BibitemShut {NoStop}%
\bibitem [{\citenamefont {Lin}\ \emph {et~al.}(2008)\citenamefont {Lin},
  \citenamefont {Lee}, \citenamefont {Kao},\ and\ \citenamefont
  {Jiang}}]{Lin2008}%
  \BibitemOpen
  \bibfield  {author} {\bibinfo {author} {\bibfnamefont {Y.}~\bibnamefont
  {Lin}}, \bibinfo {author} {\bibfnamefont {R.-K.}\ \bibnamefont {Lee}},
  \bibinfo {author} {\bibfnamefont {Y.-M.}\ \bibnamefont {Kao}}, \ and\
  \bibinfo {author} {\bibfnamefont {T.-F.}\ \bibnamefont {Jiang}},\ }\href
  {\doibase 10.1103/PhysRevA.78.023629} {\bibfield  {journal} {\bibinfo
  {journal} {Phys. Rev. A}\ }\textbf {\bibinfo {volume} {78}},\ \bibinfo
  {pages} {023629} (\bibinfo {year} {2008})}\BibitemShut {NoStop}%
\bibitem [{\citenamefont {Brazhnyi}\ and\ \citenamefont
  {Konotop}(2004)}]{Brazhnyi2004}%
  \BibitemOpen
  \bibfield  {author} {\bibinfo {author} {\bibfnamefont {V.}~\bibnamefont
  {Brazhnyi}}\ and\ \bibinfo {author} {\bibfnamefont {V.}~\bibnamefont
  {Konotop}},\ }\href {\doibase 10.1142/S0217984904007190} {\bibfield
  {journal} {\bibinfo  {journal} {Mod. Phys. Lett. B}\ }\textbf {\bibinfo
  {volume} {18}},\ \bibinfo {pages} {627 } (\bibinfo {year}
  {2004})}\BibitemShut {NoStop}%
\bibitem [{\citenamefont {Judd}\ \emph {et~al.}(2008)\citenamefont {Judd},
  \citenamefont {Scott},\ and\ \citenamefont {Fromhold}}]{Judd2008}%
  \BibitemOpen
  \bibfield  {author} {\bibinfo {author} {\bibfnamefont {T.~E.}\ \bibnamefont
  {Judd}}, \bibinfo {author} {\bibfnamefont {R.~G.}\ \bibnamefont {Scott}}, \
  and\ \bibinfo {author} {\bibfnamefont {T.~M.}\ \bibnamefont {Fromhold}},\
  }\href {\doibase 10.1103/PhysRevA.78.053623} {\bibfield  {journal} {\bibinfo
  {journal} {Phys. Rev. A}\ }\textbf {\bibinfo {volume} {78}},\ \bibinfo
  {pages} {053623} (\bibinfo {year} {2008})}\BibitemShut {NoStop}%
\bibitem [{\citenamefont {Judd}\ \emph {et~al.}(2010)\citenamefont {Judd},
  \citenamefont {Scott}, \citenamefont {Sinuco}, \citenamefont {Montgomery},
  \citenamefont {Martin}, \citenamefont {Kr\"uger},\ and\ \citenamefont
  {Fromhold}}]{Judd2010}%
  \BibitemOpen
  \bibfield  {author} {\bibinfo {author} {\bibfnamefont {T.~E.}\ \bibnamefont
  {Judd}}, \bibinfo {author} {\bibfnamefont {R.~G.}\ \bibnamefont {Scott}},
  \bibinfo {author} {\bibfnamefont {G.}~\bibnamefont {Sinuco}}, \bibinfo
  {author} {\bibfnamefont {T.~W.~A.}\ \bibnamefont {Montgomery}}, \bibinfo
  {author} {\bibfnamefont {A.~M.}\ \bibnamefont {Martin}}, \bibinfo {author}
  {\bibfnamefont {P.}~\bibnamefont {Kr\"uger}}, \ and\ \bibinfo {author}
  {\bibfnamefont {T.~M.}\ \bibnamefont {Fromhold}},\ }\href
  {http://stacks.iop.org/1367-2630/12/i=6/a=063033} {\bibfield  {journal}
  {\bibinfo  {journal} {New J. Phys.}\ }\textbf {\bibinfo {volume} {12}},\
  \bibinfo {pages} {063033} (\bibinfo {year} {2010})}\BibitemShut {NoStop}%
\bibitem [{\citenamefont {G\'oral}\ and\ \citenamefont
  {Santos}(2002)}]{Goral2002b}%
  \BibitemOpen
  \bibfield  {author} {\bibinfo {author} {\bibfnamefont {K.}~\bibnamefont
  {G\'oral}}\ and\ \bibinfo {author} {\bibfnamefont {L.}~\bibnamefont
  {Santos}},\ }\href {\doibase 10.1103/PhysRevA.66.023613} {\bibfield
  {journal} {\bibinfo  {journal} {Phys. Rev. A}\ }\textbf {\bibinfo {volume}
  {66}},\ \bibinfo {pages} {023613} (\bibinfo {year} {2002})}\BibitemShut
  {NoStop}%
\bibitem [{Note1()}]{Note1}%
  \BibitemOpen
  \bibinfo {note} {The scattering length can be tuned through a wide range
  using Feshbach resonances. This is a typical value \cite
  {Lahaye2007}.}\BibitemShut {Stop}%
\bibitem [{\citenamefont {Lahaye}\ \emph {et~al.}(2008)\citenamefont {Lahaye},
  \citenamefont {Metz}, \citenamefont {Fr\"ohlich}, \citenamefont {Koch},
  \citenamefont {Meister}, \citenamefont {Griesmaier}, \citenamefont {Pfau},
  \citenamefont {Saito}, \citenamefont {Kawaguchi},\ and\ \citenamefont
  {Ueda}}]{Lahaye2008}%
  \BibitemOpen
  \bibfield  {author} {\bibinfo {author} {\bibfnamefont {T.}~\bibnamefont
  {Lahaye}}, \bibinfo {author} {\bibfnamefont {J.}~\bibnamefont {Metz}},
  \bibinfo {author} {\bibfnamefont {B.}~\bibnamefont {Fr\"ohlich}}, \bibinfo
  {author} {\bibfnamefont {T.}~\bibnamefont {Koch}}, \bibinfo {author}
  {\bibfnamefont {M.}~\bibnamefont {Meister}}, \bibinfo {author} {\bibfnamefont
  {A.}~\bibnamefont {Griesmaier}}, \bibinfo {author} {\bibfnamefont
  {T.}~\bibnamefont {Pfau}}, \bibinfo {author} {\bibfnamefont {H.}~\bibnamefont
  {Saito}}, \bibinfo {author} {\bibfnamefont {Y.}~\bibnamefont {Kawaguchi}}, \
  and\ \bibinfo {author} {\bibfnamefont {M.}~\bibnamefont {Ueda}},\ }\href
  {\doibase 10.1103/PhysRevLett.101.080401} {\bibfield  {journal} {\bibinfo
  {journal} {Phys. Rev. Lett.}\ }\textbf {\bibinfo {volume} {101}},\ \bibinfo
  {pages} {080401} (\bibinfo {year} {2008})}\BibitemShut {NoStop}%
\bibitem [{\citenamefont {Pethick}\ and\ \citenamefont
  {Smith}(2008)}]{Pethick2008}%
  \BibitemOpen
  \bibfield  {author} {\bibinfo {author} {\bibfnamefont {C.~J.}\ \bibnamefont
  {Pethick}}\ and\ \bibinfo {author} {\bibfnamefont {H.}~\bibnamefont
  {Smith}},\ }\href
  {http://bvbr.bib-bvb.de:8991/F?func=service&doc_library=BVB01&doc_number=018677947&line_number=0001&func_code=DB_RECORDS&service_type=MEDIA}
  {\emph {\bibinfo {title} {Bose-Einstein condensation in dilute gases}}},\
  \bibinfo {edition} {2nd}\ ed.\ (\bibinfo  {publisher} {Cambridge Univ.
  Press},\ \bibinfo {address} {Cambridge},\ \bibinfo {year} {2008})\BibitemShut
  {NoStop}%
\bibitem [{\citenamefont {Penrose}\ and\ \citenamefont
  {Onsager}(1956)}]{Penrose1956}%
  \BibitemOpen
  \bibfield  {author} {\bibinfo {author} {\bibfnamefont {O.}~\bibnamefont
  {Penrose}}\ and\ \bibinfo {author} {\bibfnamefont {L.}~\bibnamefont
  {Onsager}},\ }\href {\doibase 10.1103/PhysRev.104.576} {\bibfield  {journal}
  {\bibinfo  {journal} {Phys. Rev.}\ }\textbf {\bibinfo {volume} {104}},\
  \bibinfo {pages} {576} (\bibinfo {year} {1956})}\BibitemShut {NoStop}%
\bibitem [{\citenamefont {G\'oral}\ \emph
  {et~al.}(2002{\natexlab{b}})\citenamefont {G\'oral}, \citenamefont {Gajda},\
  and\ \citenamefont {Rza\ifmmode \mbox{\c{}}\else
  \c{}\fi{}\ifmmode~\dot{z}\else \.{z}\fi{}ewski}}]{Goral2002}%
  \BibitemOpen
  \bibfield  {author} {\bibinfo {author} {\bibfnamefont {K.}~\bibnamefont
  {G\'oral}}, \bibinfo {author} {\bibfnamefont {M.}~\bibnamefont {Gajda}}, \
  and\ \bibinfo {author} {\bibfnamefont {K.}~\bibnamefont {Rza\ifmmode
  \mbox{\c{}}\else \c{}\fi{}\ifmmode~\dot{z}\else \.{z}\fi{}ewski}},\ }\href
  {\doibase 10.1103/PhysRevA.66.051602} {\bibfield  {journal} {\bibinfo
  {journal} {Phys. Rev. A}\ }\textbf {\bibinfo {volume} {66}},\ \bibinfo
  {pages} {051602} (\bibinfo {year} {2002}{\natexlab{b}})}\BibitemShut
  {NoStop}%
\bibitem [{\citenamefont {Dodd}\ \emph {et~al.}(1997)\citenamefont {Dodd},
  \citenamefont {Clark}, \citenamefont {Edwards},\ and\ \citenamefont
  {Burnett}}]{Dodd1997}%
  \BibitemOpen
  \bibfield  {author} {\bibinfo {author} {\bibfnamefont {R.}~\bibnamefont
  {Dodd}}, \bibinfo {author} {\bibfnamefont {C.}~\bibnamefont {Clark}},
  \bibinfo {author} {\bibfnamefont {M.}~\bibnamefont {Edwards}}, \ and\
  \bibinfo {author} {\bibfnamefont {K.}~\bibnamefont {Burnett}},\ }\href
  {\doibase 10.1364/OE.1.000284} {\bibfield  {journal} {\bibinfo  {journal}
  {Opt. Express}\ }\textbf {\bibinfo {volume} {10}},\ \bibinfo {pages} {284}
  (\bibinfo {year} {1997})}\BibitemShut {NoStop}%
\bibitem [{\citenamefont {Blakie}\ and\ \citenamefont
  {Davis}(2005)}]{Blakie2005}%
  \BibitemOpen
  \bibfield  {author} {\bibinfo {author} {\bibfnamefont {P.~B.}\ \bibnamefont
  {Blakie}}\ and\ \bibinfo {author} {\bibfnamefont {M.~J.}\ \bibnamefont
  {Davis}},\ }\href {\doibase 10.1103/PhysRevA.72.063608} {\bibfield  {journal}
  {\bibinfo  {journal} {Phys. Rev. A}\ }\textbf {\bibinfo {volume} {72}},\
  \bibinfo {pages} {063608} (\bibinfo {year} {2005})}\BibitemShut {NoStop}%
\bibitem [{\citenamefont {Huang}\ and\ \citenamefont {Meng}(1992)}]{Huang1992}%
  \BibitemOpen
  \bibfield  {author} {\bibinfo {author} {\bibfnamefont {K.}~\bibnamefont
  {Huang}}\ and\ \bibinfo {author} {\bibfnamefont {H.-F.}\ \bibnamefont
  {Meng}},\ }\href {\doibase 10.1103/PhysRevLett.69.644} {\bibfield  {journal}
  {\bibinfo  {journal} {Phys. Rev. Lett.}\ }\textbf {\bibinfo {volume} {69}},\
  \bibinfo {pages} {644} (\bibinfo {year} {1992})}\BibitemShut {NoStop}%
\bibitem [{\citenamefont {Krumnow}\ and\ \citenamefont
  {Pelster}(2011)}]{Krumnow2011}%
  \BibitemOpen
  \bibfield  {author} {\bibinfo {author} {\bibfnamefont {C.}~\bibnamefont
  {Krumnow}}\ and\ \bibinfo {author} {\bibfnamefont {A.}~\bibnamefont
  {Pelster}},\ }\href {\doibase 10.1103/PhysRevA.84.021608} {\bibfield
  {journal} {\bibinfo  {journal} {Phys. Rev. A}\ }\textbf {\bibinfo {volume}
  {84}},\ \bibinfo {pages} {021608} (\bibinfo {year} {2011})}\BibitemShut
  {NoStop}%
\bibitem [{\citenamefont {Saito}\ \emph {et~al.}(2012)\citenamefont {Saito},
  \citenamefont {Danshita}, \citenamefont {Ozaki},\ and\ \citenamefont
  {Nikuni}}]{Saito2012}%
  \BibitemOpen
  \bibfield  {author} {\bibinfo {author} {\bibfnamefont {T.}~\bibnamefont
  {Saito}}, \bibinfo {author} {\bibfnamefont {I.}~\bibnamefont {Danshita}},
  \bibinfo {author} {\bibfnamefont {T.}~\bibnamefont {Ozaki}}, \ and\ \bibinfo
  {author} {\bibfnamefont {T.}~\bibnamefont {Nikuni}},\ }\href {\doibase
  10.1103/PhysRevA.86.023623} {\bibfield  {journal} {\bibinfo  {journal} {Phys.
  Rev. A}\ }\textbf {\bibinfo {volume} {86}},\ \bibinfo {pages} {023623}
  (\bibinfo {year} {2012})}\BibitemShut {NoStop}%
\bibitem [{\citenamefont {Scott}\ \emph {et~al.}(2004)\citenamefont {Scott},
  \citenamefont {Martin}, \citenamefont {Bujkiewicz}, \citenamefont {Fromhold},
  \citenamefont {Malossi}, \citenamefont {Morsch}, \citenamefont {Cristiani},\
  and\ \citenamefont {Arimondo}}]{Scott2004}%
  \BibitemOpen
  \bibfield  {author} {\bibinfo {author} {\bibfnamefont {R.~G.}\ \bibnamefont
  {Scott}}, \bibinfo {author} {\bibfnamefont {A.~M.}\ \bibnamefont {Martin}},
  \bibinfo {author} {\bibfnamefont {S.}~\bibnamefont {Bujkiewicz}}, \bibinfo
  {author} {\bibfnamefont {T.~M.}\ \bibnamefont {Fromhold}}, \bibinfo {author}
  {\bibfnamefont {N.}~\bibnamefont {Malossi}}, \bibinfo {author} {\bibfnamefont
  {O.}~\bibnamefont {Morsch}}, \bibinfo {author} {\bibfnamefont
  {M.}~\bibnamefont {Cristiani}}, \ and\ \bibinfo {author} {\bibfnamefont
  {E.}~\bibnamefont {Arimondo}},\ }\href {\doibase 10.1103/PhysRevA.69.033605}
  {\bibfield  {journal} {\bibinfo  {journal} {Phys. Rev. A}\ }\textbf {\bibinfo
  {volume} {69}},\ \bibinfo {pages} {033605} (\bibinfo {year}
  {2004})}\BibitemShut {NoStop}%
\bibitem [{Note2()}]{Note2}%
  \BibitemOpen
  \bibinfo {note} {We have not performed these calculations including the
  dipolar interactions for the numerical reasons already
  discussed.}\BibitemShut {Stop}%
\bibitem [{\citenamefont {Lahaye}\ \emph {et~al.}(2007)\citenamefont {Lahaye},
  \citenamefont {Koch}, \citenamefont {Fr\"ohlich}, \citenamefont {Fattori},
  \citenamefont {Metz}, \citenamefont {Griesmaier}, \citenamefont
  {Giovanazzi},\ and\ \citenamefont {Pfau}}]{Lahaye2007}%
  \BibitemOpen
  \bibfield  {author} {\bibinfo {author} {\bibfnamefont {T.}~\bibnamefont
  {Lahaye}}, \bibinfo {author} {\bibfnamefont {T.}~\bibnamefont {Koch}},
  \bibinfo {author} {\bibfnamefont {B.}~\bibnamefont {Fr\"ohlich}}, \bibinfo
  {author} {\bibfnamefont {M.}~\bibnamefont {Fattori}}, \bibinfo {author}
  {\bibfnamefont {J.}~\bibnamefont {Metz}}, \bibinfo {author} {\bibfnamefont
  {A.}~\bibnamefont {Griesmaier}}, \bibinfo {author} {\bibfnamefont
  {S.}~\bibnamefont {Giovanazzi}}, \ and\ \bibinfo {author} {\bibfnamefont
  {T.}~\bibnamefont {Pfau}},\ }\href {\doibase doi:10.1038/nature06036}
  {\bibfield  {journal} {\bibinfo  {journal} {Nature}\ }\textbf {\bibinfo
  {volume} {448}},\ \bibinfo {pages} {672} (\bibinfo {year}
  {2007})}\BibitemShut {NoStop}%
\end{thebibliography}
\end{document}